\documentclass[fleqn]{2017SCGE}
\usepackage{multirow}
\newcommand\aap{A\&A}                
\newcommand\apj{ApJ}                 
\newcommand\mnras{MNRAS}             
\newcommand\nat{Nature}              

\newcommand{\degree}{^{\circ}}

\begin{document}

\ensubject{subject}

\ArticleType{News and Views}
\Year{2019} \Month{August} \Vol{60} \No{1} \DOI{xxx} \ArtNo{000000}
\ReceiveDate{August xx, 2019} \AcceptDate{xx xx, 2019}

\title{Advancing Pulsar Science with the FAST}

\author[1,2]{Jiguang Lu}{lujig@nao.cas.cn}%
\author[3]{Kejia Lee}{kjlee@pku.edu.cn}%
\author[3,4]{Renxin Xu}{r.x.xu@pku.edu.cn}%



\address[1]{CAS Key Laboratory of FAST, National Astronomical Observatories, Chinese Academy of Sciences, Beijing 100101, China}
\address[2]{Guizhou Radio Astronomy Observatory, Chinese Academy of Sciences, Guiyang 550025, China}
\address[3]{Kavli Institute for Astronomy and Astrophysics, Peking University, Beijing 100871, China}
\address[4]{Department of Astronomy, School of Physics, Peking University, Beijing 100871, China}


\abstract{%
The authors discuss potential remarkable achievements
for pulsar science with the FAST (pulsar monitoring, timing and
searching, as well as others related), and expect a FAST
era of pulsar science to come.}

%
%

\maketitle


\begin{multicols}{2}

As the rump left behind after an extremely gravity-induced supernova
of an evolved massive star, a pulsar is made of cool CBM (ie.,
compressed baryonic matter at low temperature).
Pulsars are not only testbed for fundamental interactions (e.g., the nature of
gravity~\cite{ShaoLJ} and of strong force at low-energy
scale~\cite{XuRX}), but also essential tools to detect nano-Hz
gravitational wave~\cite{LeeKJ}.
The pulsar science, whatever, usually depends on the measurement of
pulsar radiation, e.g., pulsar {\em monitoring} and {\em timing}.
Additionally, {\em searching} new pulsars for further investigation
is also focused in this research field.

Pulsars have a very good showing, and have never stopped presenting
surprises since the first discovery in 1967, because of advanced
facilities.
The biggest single-dish radio telescope, Chinese FAST
(Five-hundred-meter Aperture Spherical radio Telescope), is going to
observe pulsars regularly, with extremely high sensitivity but
without complicated data processing for antenna array.
The signal-to-noise ratio is $\propto A\sqrt{t}$, with $A$ the
effective area and $t$ the observing time, and a 1-minute
observation with the FAST can thus be comparable to a 10-hour action
with a 60-meter telescope if the same receiver keeps stable for such
a long time.
Therefore, we could expect a FAST era of pulsar science to come.

In this essay, we are discussing potential remarkable achievements
for pulsar science with the FAST: pulsar monitoring, timing and
searching, as well as others related.

\noindent
\begin{itemize}
\item Pulsar monitoring

\subitem The FAST with high-sensitivity can provide high signal-to-noise
ratio data, with subtle and dynamical structure information.
This means that it is not always necessary to superimpose the pulses during the analysis,
and the change in the pulse sequence between different cycles can be obtained well.
New data dimensions (e.g., phase and polarization) beyond frequency will be opened up for more information.
In fact, the correlation between multiple sets of data dimensions could
open a new window to understand pulsar magnetospheric activity~\cite{lu19a}.

\subitem Single pulse phenomena, e.g., drifting sub-pulse,
pulse nulling, mode change and giant pulse, were observed in some pulsars.
High signal-to-noise ratio data allows these phenomena to be found in more pulsars,
and new manifestations of these phenomena can be analyzed~\cite{lu19b,yu19}.
The single-pulse phenomenon is related to the physical processes
in the pulsar magnetosphere, and studying their performance and
statistical properties can help to analyze the radiation mechanism.

\subitem Pulsar is a powerful tool for studying interstellar medium.
The FAST could provide high precision data to measure the distribution
and turbulence of interstellar medium (not only free electrons, but
also the atomic gases~\cite{jene10}).
In fact, the polarization-calibrated data could also be used to determine the  interstellar magnetic fields~\cite{han94}.

\subitem The physical environment on the pulsar is extreme, with
strong electromagnetic fields, strong gravitational fields, and
supranuclear dense matter, to be difficult or intractable to
manufacture in terrestrial laboratories.
Therefore, understanding the physical processes on the pulsar may
help to discover some fundamental points.
Because of non-linearity, the physical processes in strong
electromagnetic fields are very complex and difficult to calculate with certainty.
The non-perturbative quantum chromodynamics associated
with the cool dense matter cannot even be theoretically solved.
Whereas the pulsar observation data can provide relevant information.

\item Pulsar timing

\subitem Pulsar orbital parameters (for pulsar binary/triple system) can be obtained via timing.
From high precision orbital parameters, pulsar mass can be calculated in some cases.
The pulsar mass is important for knowing the matter state and pulsar formation.
The conventional neutron star model does not allow pulsar mass to be
higher than $\sim 2.5$ or less than 0.1 times the mass of the sun,
but the strangeon star model do~\cite{XuRX}.
The mass distribution of isolate or binary pulsars can limit the pulsar formation and evolution models.

\subitem The general relativity (GR) is thought to be the standard
theory for gravity.
It is based on the assumption of the strong equivalence principle.
However, there is no ab-initial argument to support this hypothesis.
The orbits of the compact double pulsar system are seriously
deviated from the Newtonian gravitational theory, and timing on them
can effectively verify the reliability of the GR~\cite{ShaoLJ}.

\subitem Radio pulsars are rotation powered, and therefore most of
them spin down gradually.
Various braking medium are proposed, e.g., magnetic dipole
radiation, GW radiation, and stellar wind, with different spin
evolution behaviors~\cite{tong16}.
High-precision brake index measurement can help to understand the braking physics of pulsars.

\subitem In the new era of gravitational wave astronomy (GW), it is
surely urgent to open a new GW window at the low frequency ($\sim$
nHz) by the pulsar timing array (PTA) in the coming
years~\cite{fost90,hobb19}.
The European Pulsar Timing Array (EPTA), Parkes Pulsar Timing Array
(PPTA), the North American Nanohertz Observatory for GWs (NANOGrav)
are working in great efforts, and they are also collaborated to form
the International Pulsar Timing Array (IPTA) to search GW together.
While Chinese Pulsar Timing Array is being
formed\footnote{http://kiaa.pku.edu.cn/news/2017/first-chinese-pulsar-timing-array-meeting-held-kiaa},
the FAST will definitely play an important role in the competition.
In addition to the GW information veiled in the 2nd order
correlation of PTA, the 0th and the 1st order correlation could even
be applied for time standard~\cite{LiZX} and interplanetary/interstellar
navigation, and furthermore, PTA could also be used in
constructing the ephemeris of the solar system~\cite{guo18}.

\item Pulsar searching

\subitem The extremely high sensitivity of the FAST allow us to search
pulsar in the Galaxy and the nearby galaxies.
The strong pulsars in M31 is hopeful to be detected~\cite{peng00a},
which could be used to study the intergalactic medium (IGM).
With more extragalactic pulsars discovered in a galaxy, one may
understand better the galaxy's evolution.

\subitem It is encouraged to discover pulsars with extreme physics.
Pulsars with ultra high rotating speed is an indicator of pulsar's
matter state~\cite{hask18}.
It is speculated that a pulsar could be likely a strangeon star than
a neutron star if it spins with a sub-millisecond
period~\cite{DuYJ}.
Pulsars with strong magnetic fields are interesting, and their
observations may contain a wealth of strong field physics.
To discover a pulsar underneath the death-line is also important,
that means the current pulsar radiation model need to be modified.

\subitem Peculiar pulsars are also the focus of the searching.
Pulsar-black hole binary system is unpriced, which can shine in GW
theory test and pulsar population research.
If the period and its derivative of pulsar locates between the
normal pulsars and the recycled pulsars on $P-\dot{P}$ diagram, it
may provide information about pulsar evolution.
If the pulsar have large duty-cycle, it could be used to study the polarization
and magnetic field.
New single pulse phenomena may be various, and they contain information
of the radiation process and pulsar magnetosphere.

\subitem More pulsar samples help with population research.
Around 3000 pulsars have been discovered, but are just
a very small fraction of the observable pulsars in the Milky Way.
If more pulsars are discovered, then their evolution can be well
studied.
With a maximum observable zenith angle of 40$\degree$~\cite{jian19},
the FAST is expected to detect more than 4000 pulsars~\cite{li16}, to be
manifested in different forms.
The commonality of each kind of pulsars can be summarized, and the
cause of the  characteristics of each pulsar can be inferred.

\item Other fields related

\subitem The FAST can be combined with other domestic or overseas telescopes
to form a very long baseline interferometry (VLBI) network,
which can be used to accurately locate pulsars.
Measuring precisely the exact position of the pulsar in the binary
system, and together with the timing results, one can obtain the
complete information of the orbit, and consequently measure pulsar
mass.

\subitem Fast radio burst (FRB) is still puzzling now.
It has a pulse signal similar to that of pulsar, and is generally
thought to be pulsar-originated.
It is expected that FAST can observe distant FRB events and give
more detailed structure information of FRB.

\subitem The FAST can also work in the Search for Extra-Terrestrial
Intelligence (SETI) project.
Signals from distant space will be dispersed by interstellar medium,
and the readability of the signal requires some periodicity or quasi-periodicity.
Therefore, this is similar to the pulsar data, and the algorithm for
searching pulsar can be used to perform SETI related work.
Besides, pulsars are also closely related to extraterrestrial life
exploration, alien mega structures may be found around nearby
pulsars~\cite{osma18} (by the way, it is worth mentioning that the
first extrasolar planet
was found around the pulsar~\cite{wols92}).\\
\end{itemize}

In a word, the FAST can play an important role in advancing pulsar
science.
In the coming years, high-quality scientific output on pulsars will
be expected in a period of rapid development, that would certainly be
essential for us to understand the cosmic laws, from gravitational
to strong forces, etc.

\vspace*{5mm} \noindent {\small \em
This work is supported by the National Key R\&D Program of China (No. 2018YFA0404703 and 2017YFA0402602), the National Natural Science Foundation of China (Grant Nos. 11673002 and U1531243), the Strategic Priority Research Program of CAS (No. XDB23010200), and the Open Project Program of the Key Laboratory of
FAST, NAOC, Chinese Academy of Sciences.
FAST is a Chinese national mega-science facility, built and operated by the
National Astronomical Observatories, Chinese Academy of Sciences.
The FAST FELLOWSHIP is supported by Special Funding for Advanced Users, budgeted and
administrated by Center for Astronomical Mega-Science, Chinese Academy of Sciences (CAMS).}




\end{multicols}
\end{document}